\documentclass[journal]{IEEEtran}
\usepackage{etoolbox}

\ifCLASSINFOpdf
\else
\fi
\usepackage{float}

\usepackage{tabularx}
\usepackage{cite}
\usepackage[export]{adjustbox}
\usepackage{float}
\usepackage{amssymb}
\usepackage[cmex10]{amsmath}

\usepackage[caption=false,font=footnotesize]{subfig}
\usepackage{caption}
\begin{document}
	
	\title{An Accurate Analytical Approach for the Parameterization of the Single Diode Model of Photovoltaic Cell}
	\author{M.~Saqib~Ali,~\IEEEmembership{Student~Member,~IEEE,}
		and~S. M.~Raza Kazmi,~\IEEEmembership{Member,~IEEE}
\thanks{M. Saqib Ali was with the Department
	of Electrical Power Engineering, USPCAS-E, National University of Sciences and Technology (NUST), Pakistan email: saqib.qazi@ieee.org}%
\thanks{S. M. Raza Kazmi is with the Department
	of Electrical Engineering, The University of Lahore, Islamabad Campus, Pakistan email: raza.kazmi@ieee.org}}%
	\maketitle
	\begin{abstract}
	 A single diode model with five parameters is the simplest and robust approach for modeling a photovoltaic (PV) module in a simulated environment. These parameters need to be accurately extracted from the specifications given in the datasheet of the PV module such that the simulation model should exhibit the same characteristics as the actual measurements. A definite set of five independent equations, that should represent the characteristics of the PV module as accurately as possible, is needed to solve for these five parameters. In literature, the first four equations are easily created from the key data points on the characteristic curve given in the datasheet of the PV module. The main challenge however is the formulation of the fifth equation. The approaches found in literature have inherent inaccuracies due to some approximations or iterative techniques leading to discrepancy in the simulated model. This paper presents a unique analytical approach for the formulation of the fifth equation which yields the most accurate single diode model. As evident from the results, the proposed method is superior to not just the single diode model approaches but also to the double diode ones in simulating the characteristics of the PV module with least error.        
	\end{abstract}
	
	\begin{IEEEkeywords}
		Photovoltaic (PV) model, parameters extraction, analytical approach, single diode model, double diode model.
	\end{IEEEkeywords}
	
	\IEEEpeerreviewmaketitle
	\section{Introduction}
\begin{figure}[t!]
		\centering
		\subfloat[Single diode model]{%
			\includegraphics[width=3.2 in,height=1.55 in]{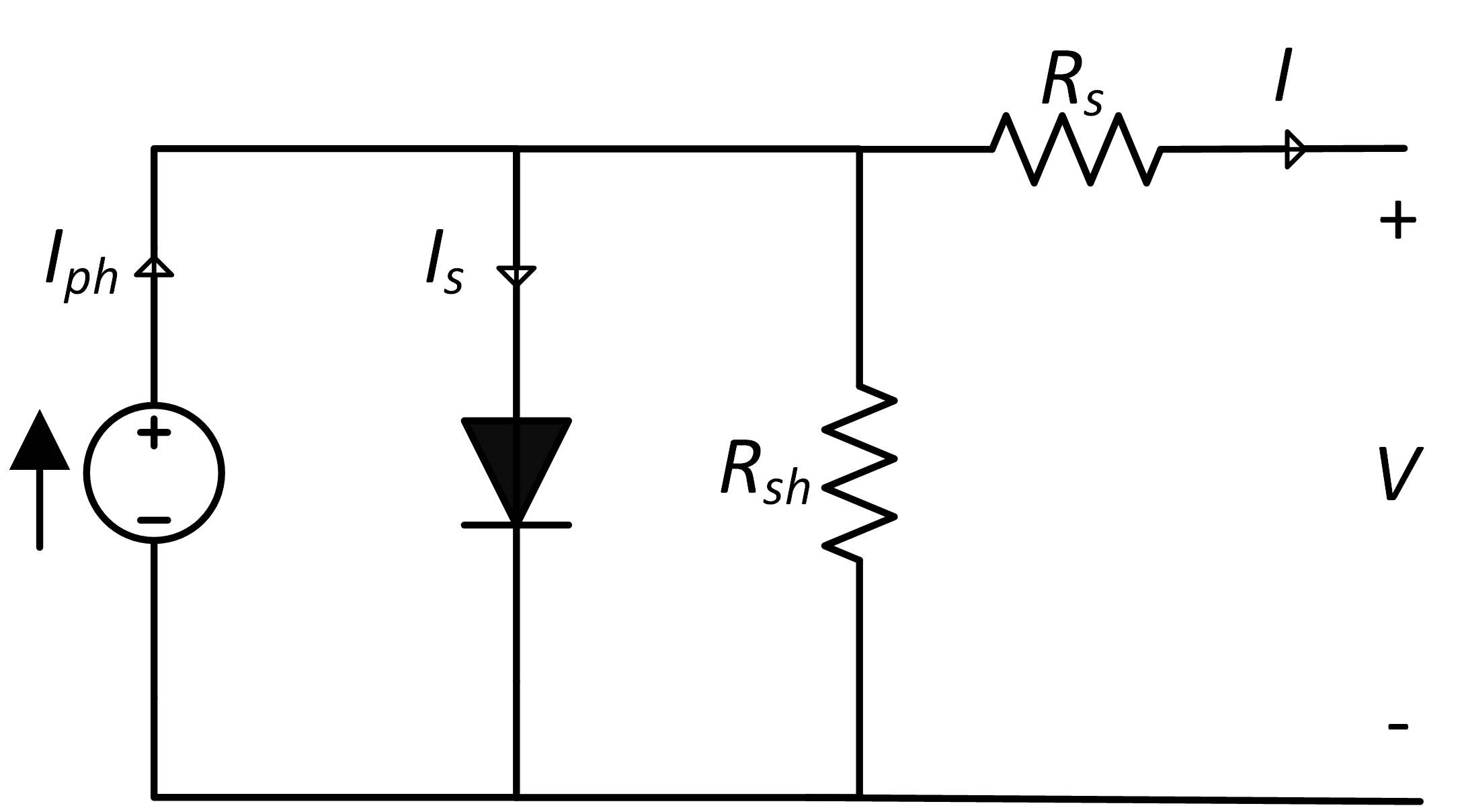}
		}\\
		\subfloat[Double diode model]{%
			\includegraphics[width=3.2 in,height=1.55 in]{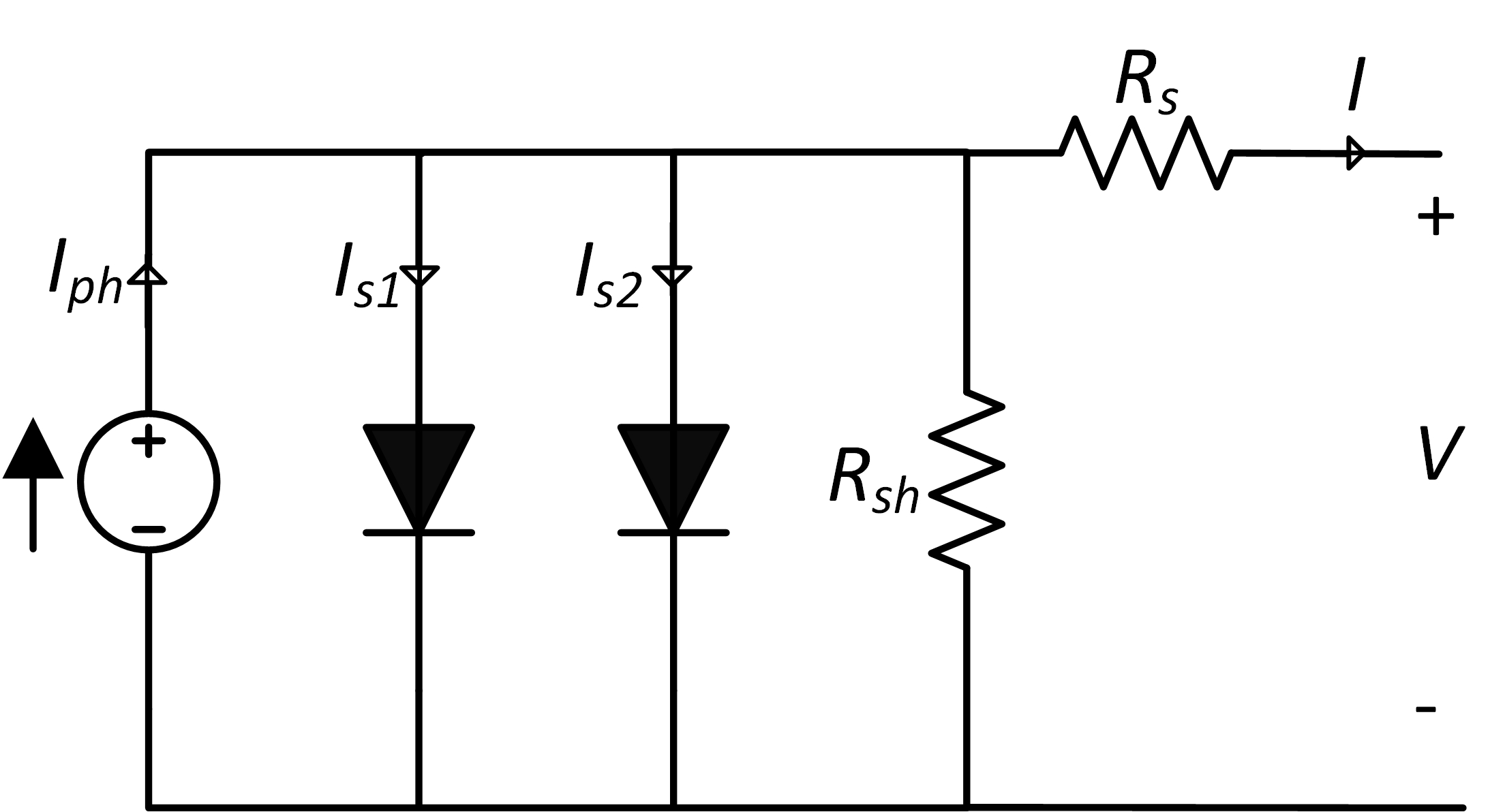}
		}%
		\captionsetup{justification=raggedright,singlelinecheck=false}	
		\caption{Equivalent circuit models of a PV module}
		\label{fig:diodes}
	\end{figure}

	\IEEEPARstart{T}{he} most important technology for the conversion of solar energy into direct electrical energy is photovoltaics. The continuous increase in efficiency of photovoltaic (PV) modules in tandem with the gradual reduction in cost and installation time, are the factors enabling solar energy to flourish rapidly \cite{energy2019international,kurtz2017new}. The planning, optimization and research for PV energy conversion system (PVECS) quite often require simulation modeling of the PV module as the fundamental building block. It is therefore of paramount significance for the equivalent simulation model to accurately mimic the actual current-voltage ($I$-$V$) and power-voltage ($P$-$V$) characteristics of the PV module. The equivalent circuits that are mostly employed in literature as simulation models of a PV module are the single diode model and the double diode model \cite{huang2015comprehensive,chan1987analytical}. As shown in Fig. \ref{fig:diodes}, the single diode model is defined by five parameters that are photon current $I_{ph}$, series and shunt resistances $R_s$ and $R_{sh}$ respectively, and the two parameters of the diode that are saturation current $I_s$ and ideality factor $n$. The double diode model has two saturation currents $I_{s1}$ and $I_{s2}$ as well as two ideality factors $n_1$ and $n_2$ due to the additional diode. $N_s$ is a known constant since it is the number of PV cells connected in series in a module. The accuracy of these models heavily depends on their parameters which are not directly extractable from the specifications in the datasheet or the measured data of the PV module.
	\par In literature various approaches for parameters extraction of a PV module are available, each strive to minimize the error between the simulated and the actual characteristic curves. Extensive reviews on parameters extraction through different models and techniques can be found in \cite{yang2020comprehensive,shongwe2015comparative,batzelis2019non,abbassi2018identification,gomes2016shuffled}. These techniques generally bifurcate into two major categories: a) searching algorithms based and b) analytical model based. The former category uses meta-heuristics and iterative search algorithms to find the best fitting parameters; while the techniques in the latter category formulate simultaneous independent equations that can be uniquely solved to find the unknown parameters.
	\subsection{Searching Algorithms Based}
	It is conspicuous from the implicit nature of current ($I$) with respect to voltage ($V$) in the equations shown in Fig. \ref{fig:diodes} that the closed form analytical solutions won't be readily available. Therefore in literature there are several iterative methods and search based algorithms to represent the current as function of voltage \cite{waly2019parameters,subudhi2017bacterial,soon2012photovoltaic,optimizing15,paviet2016accurate,moshksar2016adaptive,mathew2017wind,ibrahim2019adaptive,hasanien2015shuffled,gomes2016shuffled,oliva2017parameter,chin2017accurate,bharadwaj2016sequential,cardenas2016experimental,jadli2017new,ma2019data,elazab2020parameter}. In \cite{soon2012photovoltaic}, $I_{ph}$ and $I_s$ are predefined while $n$, $R_s$ and $R_{sh}$ are estimated. The author employed PSO with constraints on objective function to heavily penalize when the solution tries to move away from the constraints. The same authors then proposed in \cite{optimizing15} a generalized model with an array of diodes that can be added in series as well as in parallel. It is shown that introducing diodes in series increases the coverage region of the $I$-$V$ and $P$-$V$ characteristic curves of the PV model. Parameters are extracted through the similar technique presented in \cite{soon2012photovoltaic}. The \cite{paviet2016accurate} showed that the error in the estimated curve will increases with increase in fill factor. The author further proposed a fit range for more accurate estimation of peak power. The \cite{moshksar2016adaptive}, initially produced an unconstrained optimization problem to penalize the constraints and then, the gradient descent technique is applied to manage the optimal values of parameters for reduced diode model. The \cite{subudhi2017bacterial} used bacterial foraging techniques by defining a fitness function and updating it in iterations. 
	
	The \cite{mathew2017wind,ibrahim2019adaptive} proposed wind driven optimization (WDO) algorithms for extraction of parameters. The author in \cite{mathew2017wind} compares WDO technique with the several other optimization and searching techniques like PSO, GA, bee colony, flower pollination etc., and recommended WDO as the faster and much accurate algorithm. The \cite{ibrahim2019adaptive} proposed adaptive-WDO (AWDO) algorithm to have the system less reliant on user for input parameters. The \cite{oliva2017parameter} applied chaotic whale optimization (CWO) algorithm on single and double diode models. The functioning of whale optimization (WO) is improved with chaotic Singer map, which is used for generation of chaotic sequence for updating whale position in each iteration, to obtain the best set of extracted parameters. The \cite{chin2017accurate} proposed a hybrid technique for a double diode model with seven parameters; four are extracted from the analytical equation and remaining three from differential evolution (DE) that leads to a single best solution. The \cite{bharadwaj2016sequential,jadli2017new} also used amalgamation of analytical and searching optimization while \cite{cardenas2016experimental} using the reduced search space to lessen the searching effort. The \cite{ma2019data} proposed a data driven approach for the estimation of $I$-$V$ curve and for extracting parameters. Although this technique is robust however the authors have concluded that prediction of $R_s$ has predictable error.  The \cite{waly2019parameters} compared genetic algorithms (GA) with Newton Raphson (NR) and particle swarm optimization (PSO), and exhibited GA supremacy in terms of convergence and less number of iterations. The \cite{hasanien2015shuffled} used social behaviour of frogs for parameterization of a single diode model while \cite{elazab2020parameter} used  grasshopper optimization algorithm (GOA) on a three diode model and proved its accuracy compared to other meta-heuristic techniques.
	
	These searching and numerical based techniques exhibits several drawbacks despite their accuracy, enumerated as follows \cite{yang2020comprehensive,jadli2017new,ishaque2011critical,chin2017accurate}: (i) these techniques require data points of the entire characteristics curve which are not always readily available  (ii) these are highly complex and non-generic algorithms (iii) these techniques, meta-heuristics in particular, are slower due to point-by-point comparison for curve fitting (iv) GA and differential evolution (DE) based algorithms are not much effective for double diode model with seven parameters. The reason being that the values of two reverse saturation currents come out to be so close to each other that effectively the two diode model works like a single diode model \cite{lim2015linear}.
	\begin{figure}[t!]
		\centering
		\includegraphics[width=3.6 in,height=2.75 in]{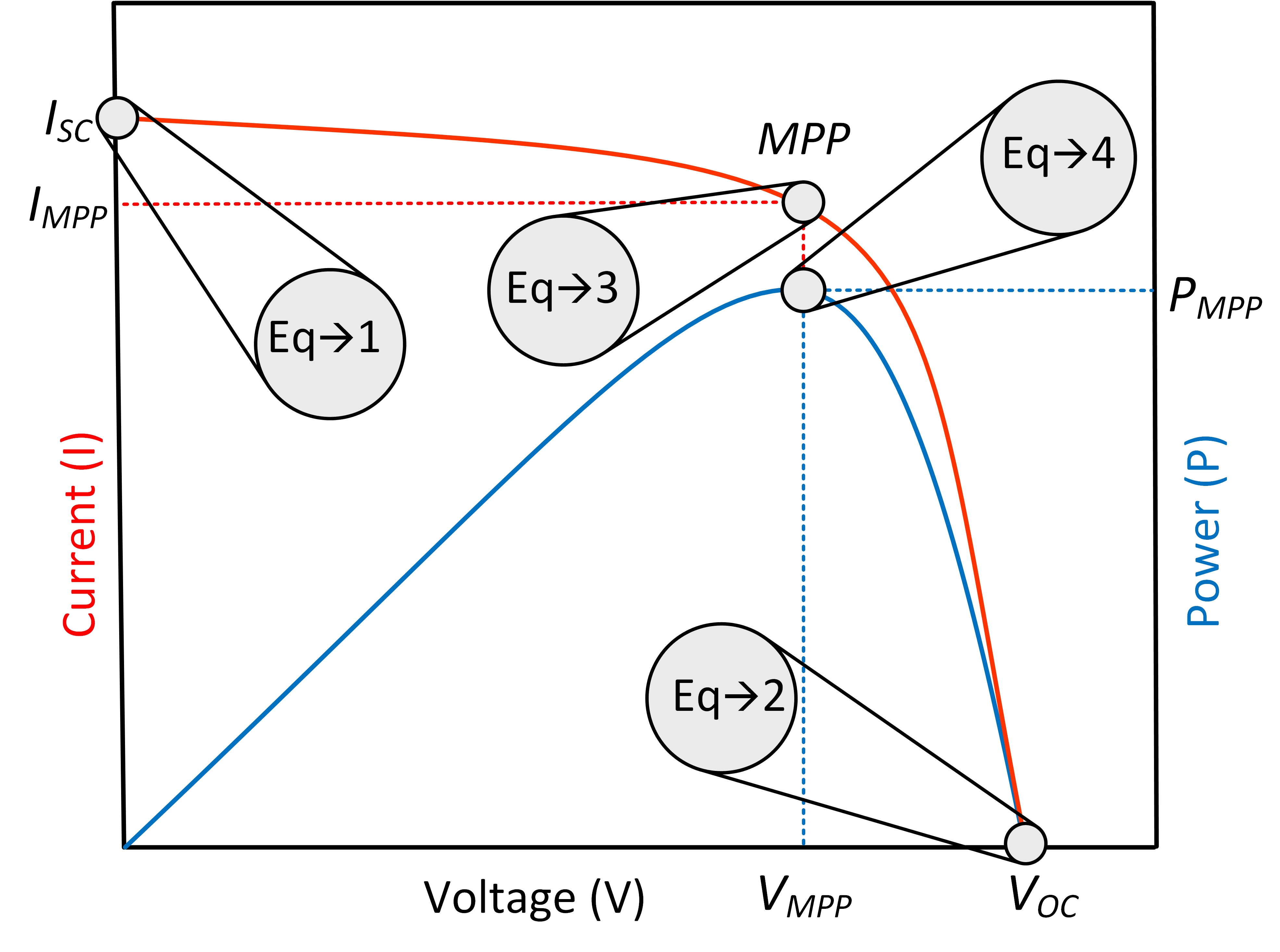}
		\captionsetup{justification=raggedright,singlelinecheck=false}
		\caption{PV module $I$-$V$ and $P$-$V$ characteristic curves and key data points for formation of first four equations}
		\label{fig:curve}
	\end{figure}  
	\subsection{Analytical Modeled Based}
	The \cite{mahmoud2012parameterization,chakrasali2013network,khezzar2014modeling,benavides2008modeling,tan2004model,saloux2011explicit,huang2015comprehensive,babu2014novel,villalva2009comprehensive,hejri2014parameter,bastidas2017genetic,lineykin2014issues,silva2015parameter,kareem2016new,sera2007pv,brano2010improved,ding2014simplified,bai2014development,arab2004loss,de2006improvement,tian2012cell,lun2013explicit,chenni2007detailed,ma2014development,laudani2014identification,batzelis2015method,mehta2019accurate} covers the analytical modeling based techniques, each proposing a set of independent equations that can be simultaneously solved for parameters extraction. Invariably in all these publications, four nonlinear analytical equations can be formed from the key data points on the characteristics curves that are shown in Fig. \ref{fig:curve} \cite{mahmoud2012parameterization}. These key data points can be found directly from the datasheet or the nameplate of PV module provided by the manufacturer. However the main issue addressed in these publications is the formulation of the additional equations required to complete the set; for instance one and three additional equations for single and double diode models respectively to match the number of unknown parameters. This is a challenging tasks for the reason that all the key data points available in datasheet have already been exhausted for the first four equations. Some researchers have circumvented the need for additional equations by: a) reducing number of parameters b) making some assumptions c) using predefined values of some parameters. For a single diode model with five parameters, the \cite{chakrasali2013network,khezzar2014modeling} neglected the $R_{sh}$ while \cite{benavides2008modeling,tan2004model} neglected the $R_s$ and reduced their models to four parameters. The \cite{saloux2011explicit} neglected both $R_s$ and $R_{sh}$ and proposed an ideal single diode model. As a result these approaches \cite{chakrasali2013network,khezzar2014modeling,benavides2008modeling,tan2004model,saloux2011explicit} suffer digression specifically at maximum power point (MPP) due to the absence of $R_s$ and $R_{sh}$ \cite{huang2015comprehensive}. The \cite{babu2014novel} proposed a double diode model and neglected $R_s$ and $R_{sh}$ and therefore the fitting precision of $I$-$V$ characteristics is highly affected. The \cite{villalva2009comprehensive} assumed the photon current $I_{ph}$ equal to the short circuit current $I_{sc}$ and fixed the value of diode ideality factor $n$. The \cite{hejri2014parameter} used double diode model and extracted parameters by tedious mathematical modeling but with some assumptions which ultimately lead to discrepancy in model accuracy.\\
	\indent The \cite{mahmoud2012parameterization} initially estimated the viable range of $R_{sh}$ separately and then extracted other four parameters from the analytical equations at each value of $R_{sh}$, and eventually selected the best values that yielded least error. However in \cite{bastidas2017genetic} it is argued that the $I$-$V$ and $P$-$V$ characteristics curves are not quite sensitive to the change in $R_{sh}$, therefore sweeping $R_{sh}$ to find other parameters may not land at accurate model \cite{moshksar2016adaptive}. The \cite{lineykin2014issues} extracted the five parameters for single diode by sweeping the $n$. After getting multiple characteristics curves with different parameters, author proposed an optimal value of $n$ with minimum error. The \cite{silva2015parameter} scanned the $n$ and $R_s$ in small steps and extracted other parameters from the analytical equations. The estimated parameters are bounded by different minimum and maximum values for different PV modules. Any kind of change in situation will surely effects these boundaries and it will ultimately increases the algorithm complexity. On the other hand, convergence problems arise from incorrect initial guesses \cite{hejri2014parameter}. Due to inadequate number of readily available equations, different researchers made some assumptions for the simplification of computational effort. These may converge to a solution but the accuracy of the solution is not guaranteed. For instance, a double diode model with equal value of two saturation currents is effectively a single diode model \cite{chin2017accurate}.\\
	\indent On the other hand several researchers have derived the fifth equation for the five parameters of single diode model \cite{kareem2016new,sera2007pv,brano2010improved,ding2014simplified,bai2014development,arab2004loss,de2006improvement,tian2012cell,lun2013explicit,chenni2007detailed,ma2014development,laudani2014identification,batzelis2015method,mehta2019accurate}. These publications till date either rely on some assumptions which lead to inaccuracies, or devise some complex techniques leading to longer turnaround time in computation. Section III presents the variants of fifth equation found in literature in contrast with the proposed technique.\\
	\indent This paper thus proposes a non-complex and accurate approach for the formulation of the fifth equation for the parameter extraction of a single diode model of PV module. Section II illustrates the mathematical modeling of PV module in terms of formulation of equations which are basically derived from its characteristic curves. Section III describes previous methods in the literature for the formation of fifth equation and explains the proposed fifth equation. Section IV presents the results which clearly establish the supremacy of the proposed technique not only against the single- but also double-diode model techniques.
		\begin{figure}[t!]
		\centering
		\includegraphics[width=3.3 in,height=2.75 in]{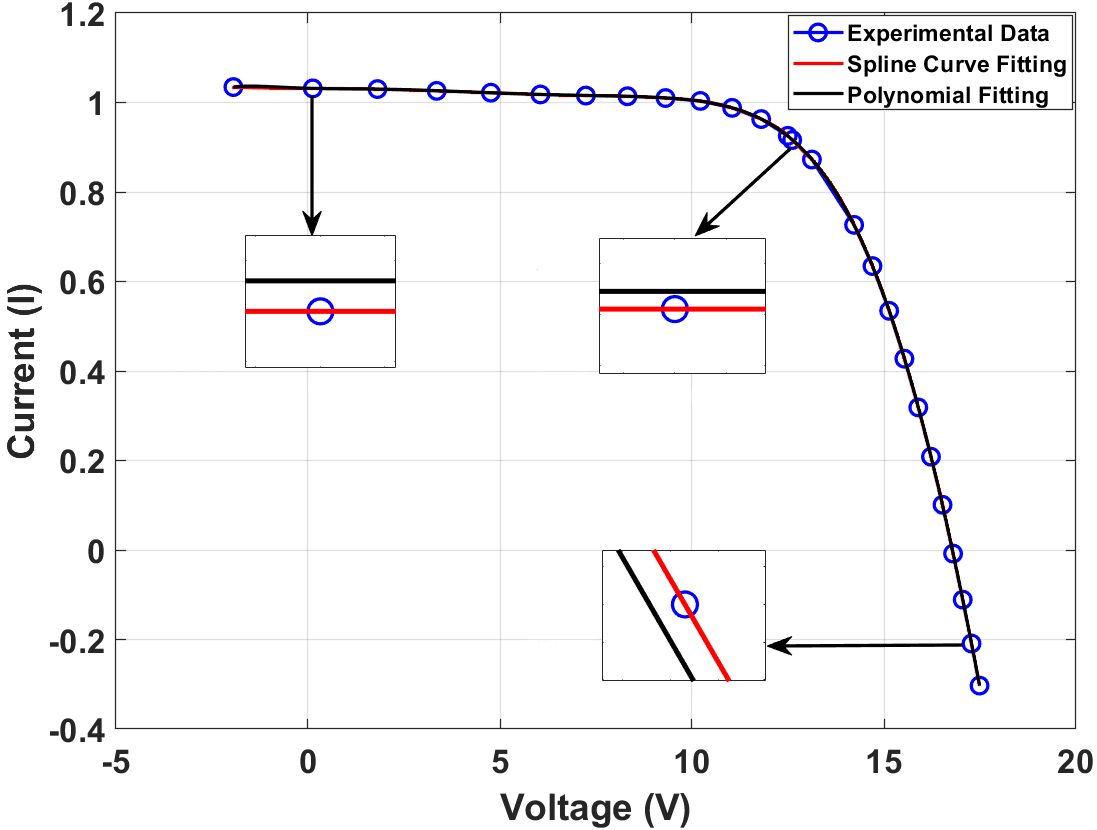}
		\captionsetup{justification=raggedright,singlelinecheck=false}
		\caption{Comparison of curve fitting techniques for PV module PWP-201}
		\label{fig:Spline}
	\end{figure}
	\section{Mathematical Modeling}  
	Equation (\ref{eq:1}) represents the current ($I$) and voltage ($V$) relationship for the single diode model of a PV module.
	\begin{equation}
	I=I_{ph}-I_s[\exp(\frac{V+IR_s}{N_s.n.v_t})-1]-\frac{V+IR_s}{R_{sh}}
	\label{eq:1}
	\end{equation}
	where the thermal voltage 
	\begin{equation*}
	v_t=\frac{kT}{q}
	\label{eq:2}
	\end{equation*}
	and\\
	\indent $q$ \hspace{10mm}Unit charge (C)\\
	\indent $T$ \hspace{9mm}Temperature of PV module (K)\\
	\indent $k$ \hspace{9.8mm}Boltzman constant (J/K)\\
	\indent $I_{ph}$ \hspace{7.2mm}Photon current (A)\\
	\indent $I_s$ \hspace{8.8mm}Reverse saturation current (A)\\
	\indent $R_s$ \hspace{7.8mm}Series resistance ($\Omega$)\\
	\indent $R_{sh}$ \hspace{6mm}Shunt resistance ($\Omega$)\\
	\indent $N_s$ \hspace{7.5mm}Number of series connected PV cells\\
	\indent $n$ \hspace{10mm}Ideality factor of diode\\
	\\
	Each PV module has specific key data points, as shown in Fig. \ref{fig:curve}. These data points, enlisted below, provide the boundary values of (1).
	\\
	\indent $I_{sc}$ \hspace{7.2mm} Short circuit current (A)\\
	\indent $V_{oc}$ \hspace{6.8mm} Open circuit voltage (V)\\
	\indent $I_{MPP}$ \hspace{2.6mm} Current at MPP (A)\\
	\indent $V_{MPP}$ \hspace{2.6 mm} Voltage at MPP (V)\\
	\indent $P_{MPP}$ \hspace{2.55mm} Power at MPP (W)\\
	\\
	These data points are the key to derive the first four equations of PV model \cite{de2006improvement, sera2007pv,laudani2014identification,lineykin2014issues}. The ($I$, $V$) in (\ref{eq:1}) are eliminated by ($I_{sc}$, 0), (0, $V_{oc}$) and ($I_{MPP}$, $V_{MPP}$) to get the equations at short circuit (SC) point, open circuit (OC) point and MPP respectively.
\begin{figure*}[t!]
	\centering
	\subfloat[Three from $I$-$V$ curve]{%
		\includegraphics[width=2.35 in,height=1.9 in]{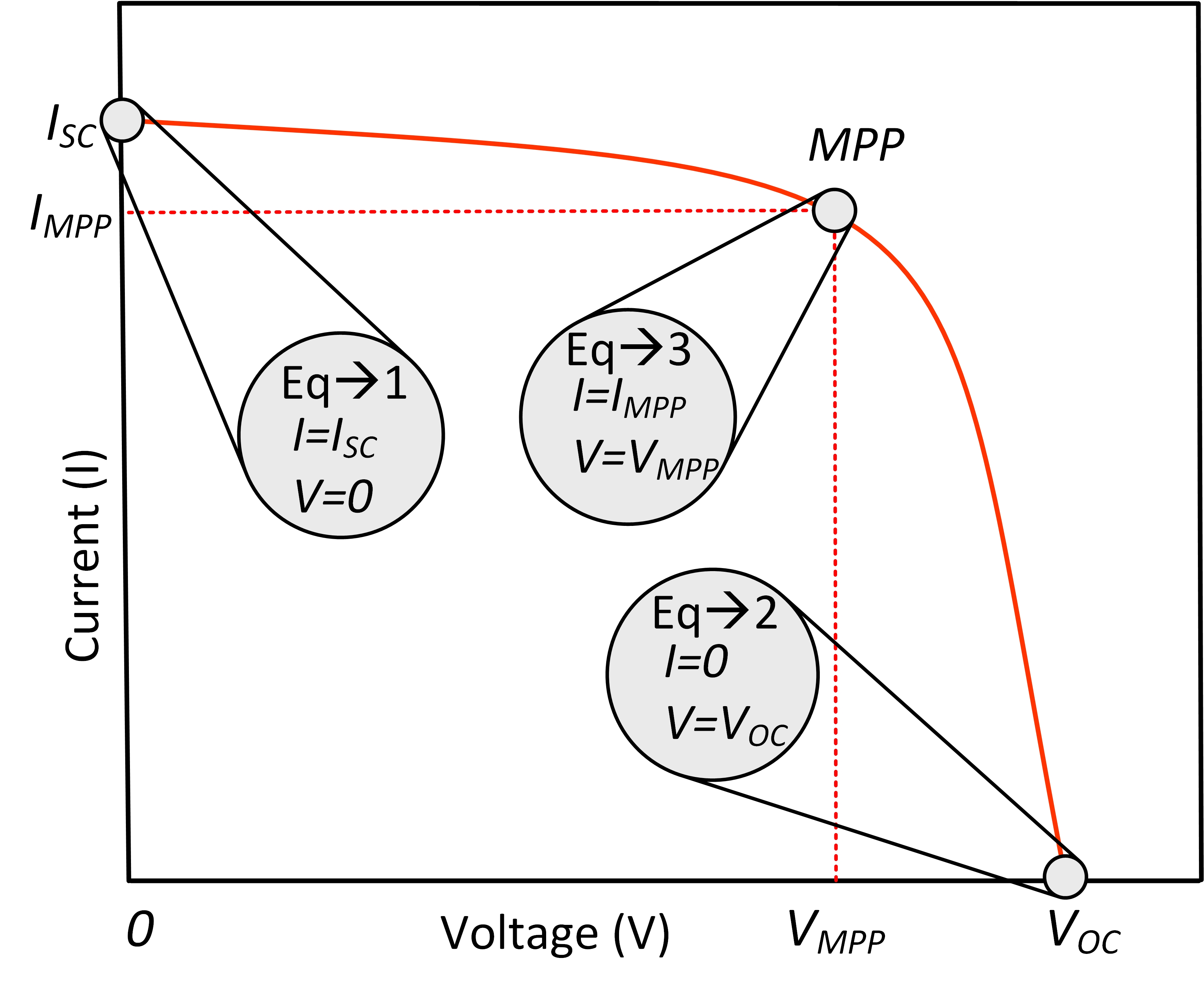}
	}%
	\subfloat[Fourth form $P$-$V$ curve]{%
		\includegraphics[width=2.35 in,height=1.9 in]{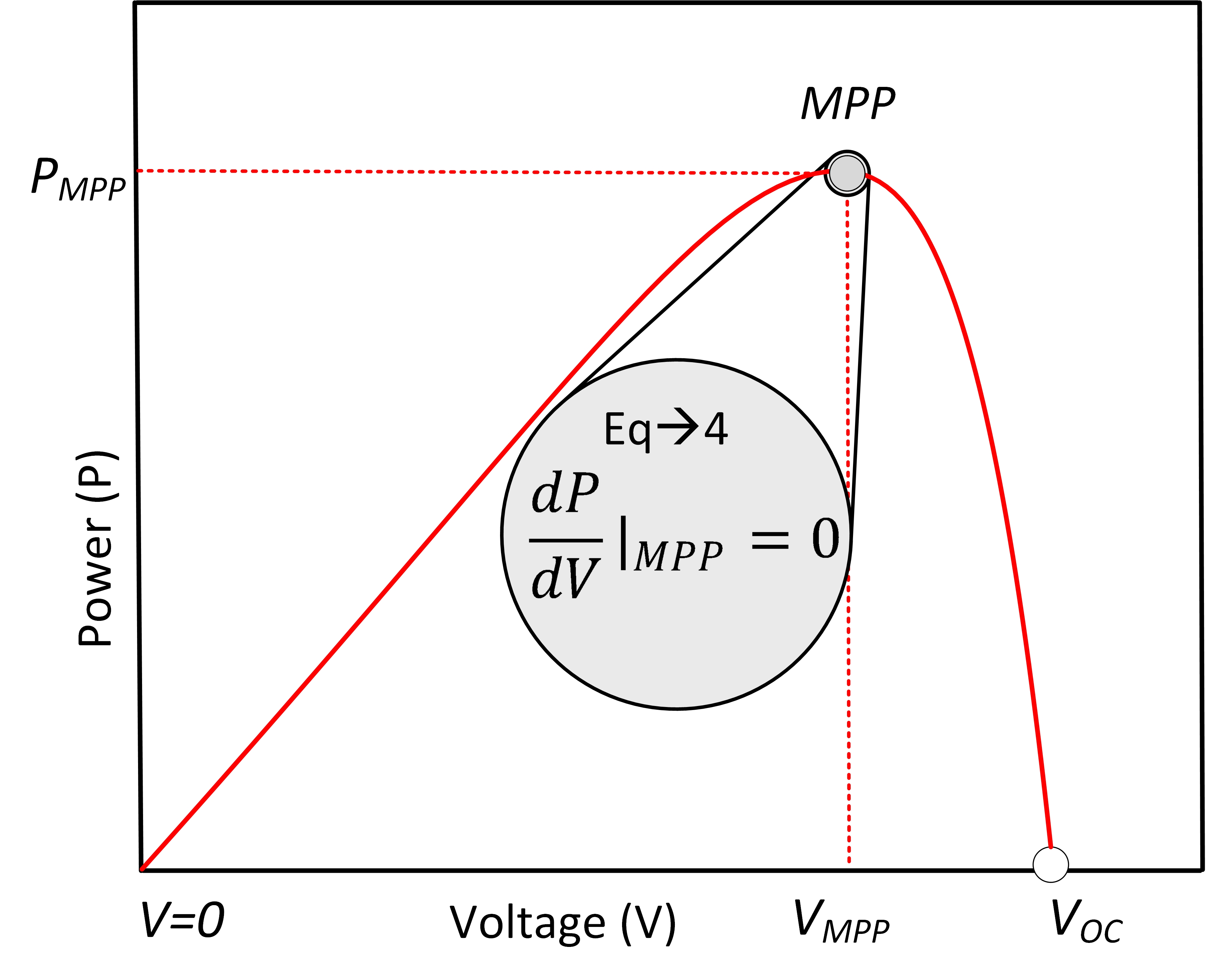}
	}
	\subfloat[Fifth from $P$-$I$ curve]{%
		\includegraphics[width=2.35 in,height=1.9 in]{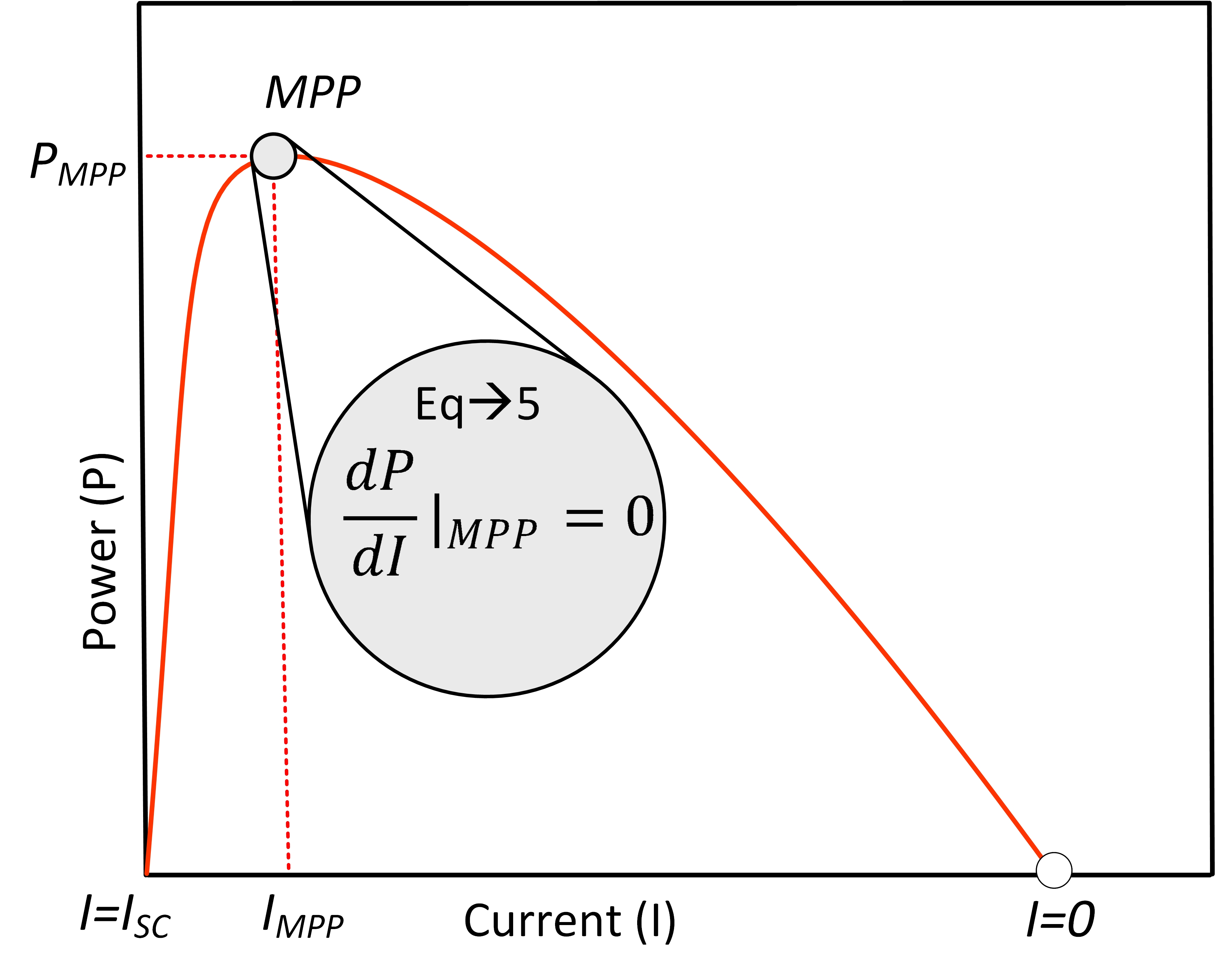}
	}%
	\captionsetup{justification=raggedright,singlelinecheck=false}	
	\caption{Overview of formation of equations form the characteristic curves of the PV module}
	\label{fig:overview} 
\end{figure*}
	\begin{equation}
	SC:I_{sc}=I_{ph}-I_s[\exp(\frac{I_{sc}R_s}{N_s.n.v_t})-1]-\frac{V+I_{sc}R_s}{R_{sh}}
	\label{eq:3}
	\end{equation}
	\begin{equation}
	OC:0=I_{ph}-I_s[\exp(\frac{V_{oc}}{N_s.n.v_t})-1]-\frac{V_{oc}}{R_{sh}}
	\label{eq:4}
	\end{equation}
	\begin{equation}
	MPP:I_{MPP}=I_{ph}-I_s[\exp(\frac{I_{sc}R_s}{N_s.n.v_t})-1]-\frac{V_{MPP}+I_{MPP}R_s}{R_{sh}}
	\label{eq:5}
	\end{equation}
	\\
	While above three equation are from the $I$-$V$ curve, the fourth equation is generated at the MPP on the $P$-$V$ characteristic curve \cite{de2006improvement, sera2007pv,laudani2014identification} using the fact that maxima is defined as
	\begin{equation}
	\left.\frac{dP}{dV}\right\vert_{{V=V_{MPP}}\atop{I=I_{MPP}}}=0	
	\label{eq:6}
	\end{equation}
	Now
	\begin{equation}
	\frac{dP}{dV}=\frac{d(IV)}{dV}=I+\frac{dI}{dV}V
	\label{eq:7}
	\end{equation}
	The (\ref{eq:1}) is a transcendental equation in the form of $I=f(V,I)$, hence its implicit differential is expressed using chain rule.
	\begin{equation}
	dI=dI\frac{\partial f(I,V)}{\partial I}+dV\frac{\partial f(I,V)}{\partial V}
	\label{eq:8}
	\end{equation}
	So, the derivative of (\ref{eq:8}) with respect to voltage is
	\begin{equation}
	\frac{dI}{dV}=\frac{\frac{\partial}{\partial V}f(I,V)}{1-\frac{\partial}{\partial I}f(I,V)}
	\label{eq:9}
	\end{equation}
	Substituing (\ref{eq:9}) in (\ref{eq:7})
	\begin{equation}
	\frac{dP}{dV}=I+V\frac{\frac{\partial}{\partial V}f(I,V)}{1-\frac{\partial}{\partial I}f(I,V)}
	\label{eq:10}
	\end{equation}
	Evaluating (\ref{eq:10}) at MPP gives the final form of the fourth equation as
	\begin{equation}
	\begin{split}
	\left.\frac{dP}{dV}\right\vert_{{V=V_{MPP}}\atop{I=I_{MPP}}}= 0 = I_{MPP}-V_{MPP}\ \times\\\frac{\frac{I_s}{N_s.n.v_t}[\exp(\frac{V_{MPP}+I_{MPP}R_s}{N_s.n.v_t})-1]+\frac{1}{R_{sh}}}{1+\frac{I_s.R_s}{N_s.n.v_t}[\exp(\frac{V_{MPP}+I_{MPP}R_s}{N_s.n.v_t})-1]+\frac{R_s}{R_{sh}}}
	\label{eq:11}
	\end{split}
	\end{equation}
	\begin{center}
		\begin{table}[]
			\caption{Specifications of PWP-201 and RTC France}
			\centering
			\begin{tabular}{ccc}
				\hline \hline
				\textbf{Parameters} & \textbf{PWP-201} & \textbf{RTC France}  \\
				\\
				\hline \hline
				 $I_{sc}$(A)& 1.03163  & 0.7603    \\
				 $V_{oc}$(V) & 16.7753 & 0.5728   \\
				 $I_{MPP}$(A) & 0.9162  & 0.6894   \\
				 $V_{MPP}$(V) & 12.6049  & 0.4507   \\
				$P_{MPP}$(W)  &  11.55  & 0.3107    \\
				\hline \hline
			\end{tabular}%
			\label{tab:parameters}%
		\end{table}%
	\end{center}
	\section{Formulation of Fifth Equation}
	From Section II it can be concluded that the derivation of the four independent equations (\ref{eq:3}), (\ref{eq:4}), (\ref{eq:5}) and (\ref{eq:11}), has exhausted all the key data points shown in Fig. \ref{fig:curve}. Therefore the formulation of the fifth equation is a non-trivial task. Variants of the fifth equation derived in previous approaches are described in the following subsections and the equation proposed in this manuscript is presented at the end.
	\subsection{Previous Approaches} 
	In \cite{kareem2016new,sera2007pv,brano2010improved,ding2014simplified,bai2014development}, the $R_{sh}$ is assumed as the slope of $I$-$V$ curve at SC point. 
	\begin{equation}
	\left.\frac{dI}{dV}\right\vert_{{I=I_{sc}}}=-\frac{1}{R_{sh}}
	\label{eq:12}
	\end{equation}  
	As a dual of (\ref{eq:12}) the slope at OC point on the $I$-$V$ curve is equated to the $R_{s}$ in \cite{brano2010improved,arab2004loss}.
	 \begin{equation}
	 \left.\frac{dI}{dV}\right\vert_{{V=V_{oc}}}=-\frac{1}{R_{s}}
	 \label{eq:13}
	 \end{equation} 
	The \cite{de2006improvement,tian2012cell,lun2013explicit,chenni2007detailed, ma2014development, batzelis2015method} make use of the temperature coefficients available in the datasheet of PV module. The constant coefficients imply that the voltage and current vary linearly with temperature. Thus the fifth equation is same as (\ref{eq:4}) but computed at an arbitrarily different temperature than standard test condition.
	
	In \cite{mehta2019accurate} the fifth equation is produced by computing the area ($A$) under the measured $I$-$V$ characteristic curve of particular PV modules, and then relating $A$ to (\ref{eq:1}) using trapezoidal method of numerical integration. It is given as
	\begin{equation}
	 A=-\frac{h.I{sc}}{2}+h\sum_{i=1}^{i=N}I_i
	 \label{eq:mehta}
	 \end{equation}
	where $h$ is the step size that divides the voltage axis into $N$ data points from the SC to the OC points and $i$ is the iteration number to compute the current $I_i$ at each data point.
	
	All of the above approaches have inherent inaccuracies. PV module datasheet never reports the slopes of $I$-$V$ characteristic curves at SC and OC points; moreover representing these slopes as (\ref{eq:12}) and (\ref{eq:13}) respectively is mere approximation usually derived through polynomial curve fitting \cite{laudani2014identification}. Similarly the linear relation of PV voltage and current with temperature is also an approximation due to the well-known nonlinear dependence of thermally generated carriers on temperature in semiconductors. The computation of area in \cite{mehta2019accurate} is done by first converting the experimental data into continuous $I$-$V$ curve through ninth order polynomial fit. Fig. \ref{fig:Spline} shows the discrepancy in this curve fitting leading to inaccurate calculation of the area. It further demonstrates that a better choice would've been the $Spline$ function of MATLAB that yields a curve fit more accurately passing through the experimental data points. Moreover the computation of (\ref{eq:mehta}) requires fine granularity for accuracy. So \cite{mehta2019accurate} had to use one millions data points to calculate the area, thus requiring larger processing time and memory.
	
	 \begin{figure}[!t]
	 	\centering
	 	\subfloat[Current-Voltage ($I$-$V$) curve]{%
	 		\includegraphics[width= 3.5 in, height= 2.55 in]{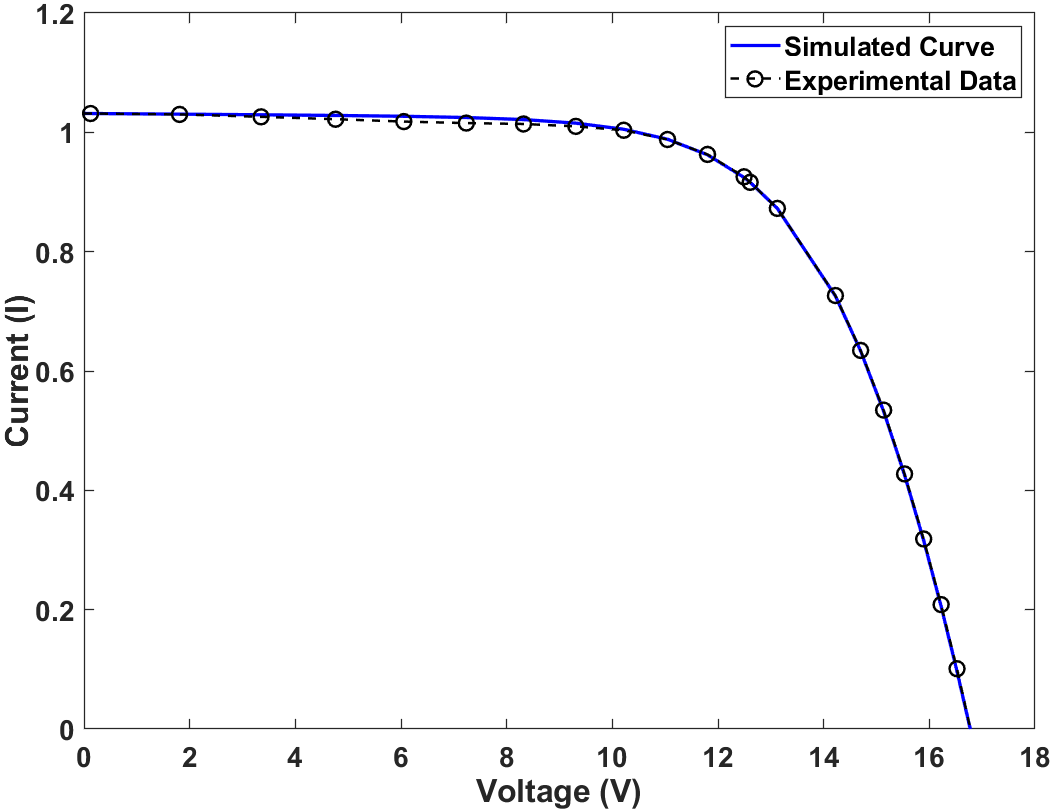}
	 	}\\
	 	\subfloat[Power-Voltage ($P$-$V$) curve]{%
	 		\includegraphics[width= 3.5 in, height= 2.55 in]{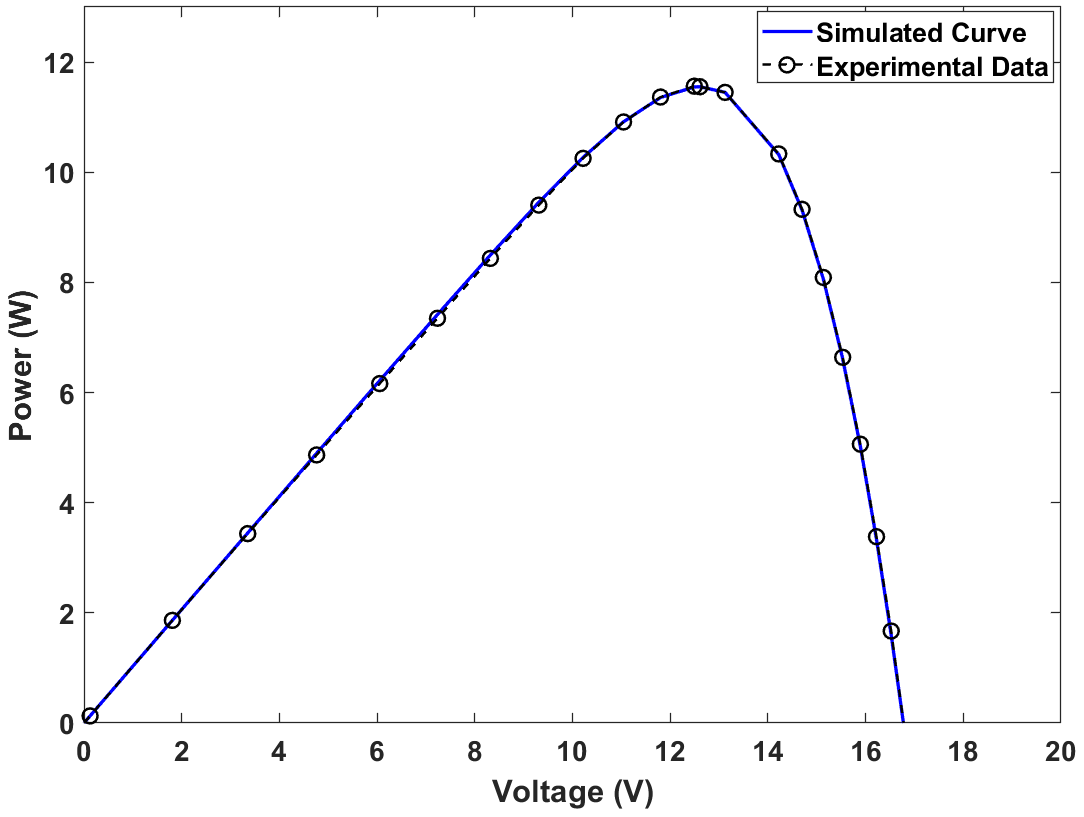}
	 	}%
	 	\captionsetup{justification=raggedright,singlelinecheck=false}	
	 	\caption{Simulated and experimental curves of PV module PWP-201}
	 	\label{fig:PWP_IV}
	 \end{figure}
	 \begin{figure}[!t]
	 	\centering
	 	\includegraphics[width= 3.5 in, height= 2.8 in]{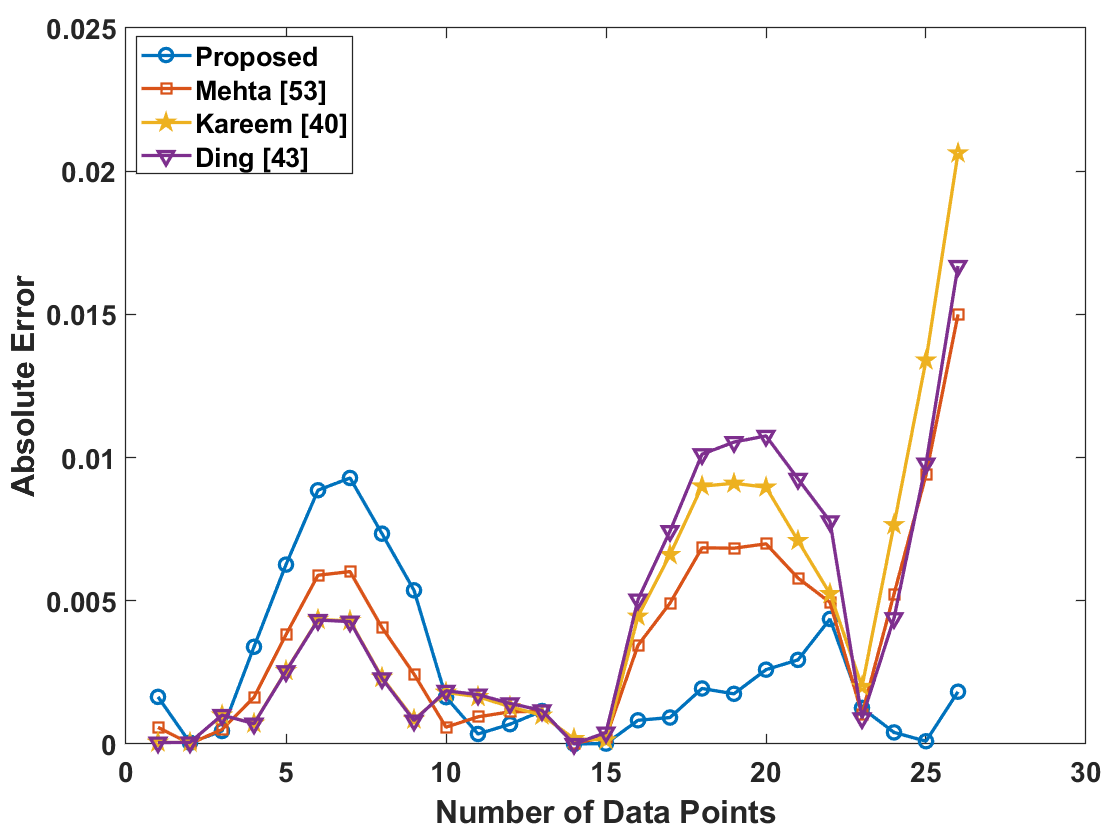}
	 	\captionsetup{justification=raggedright,singlelinecheck=false}
	 	\caption{Comparison of absolute errors for PV module PWP-201}
	 	
	 	\label{fig:Error_PWP201}
	 \end{figure}
	 \begin{figure}[!t]
	 	\centering
	 	\subfloat[Current-Voltage ($I$-$V$) curve]{%
	 		\includegraphics[width= 3.5 in, height= 2.55 in]{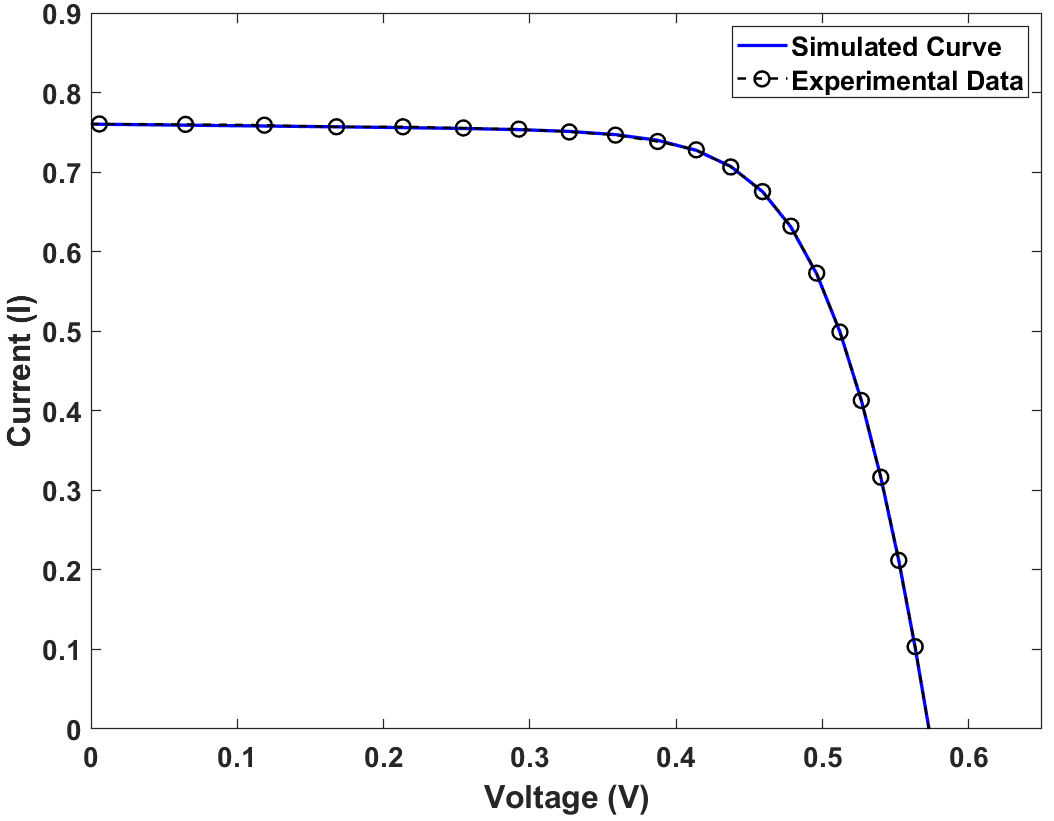}
	 	}\\
	 	\subfloat[Power-Voltage ($P$-$V$) curve]{%
	 		\includegraphics[width= 3.5 in, height= 2.55 in]{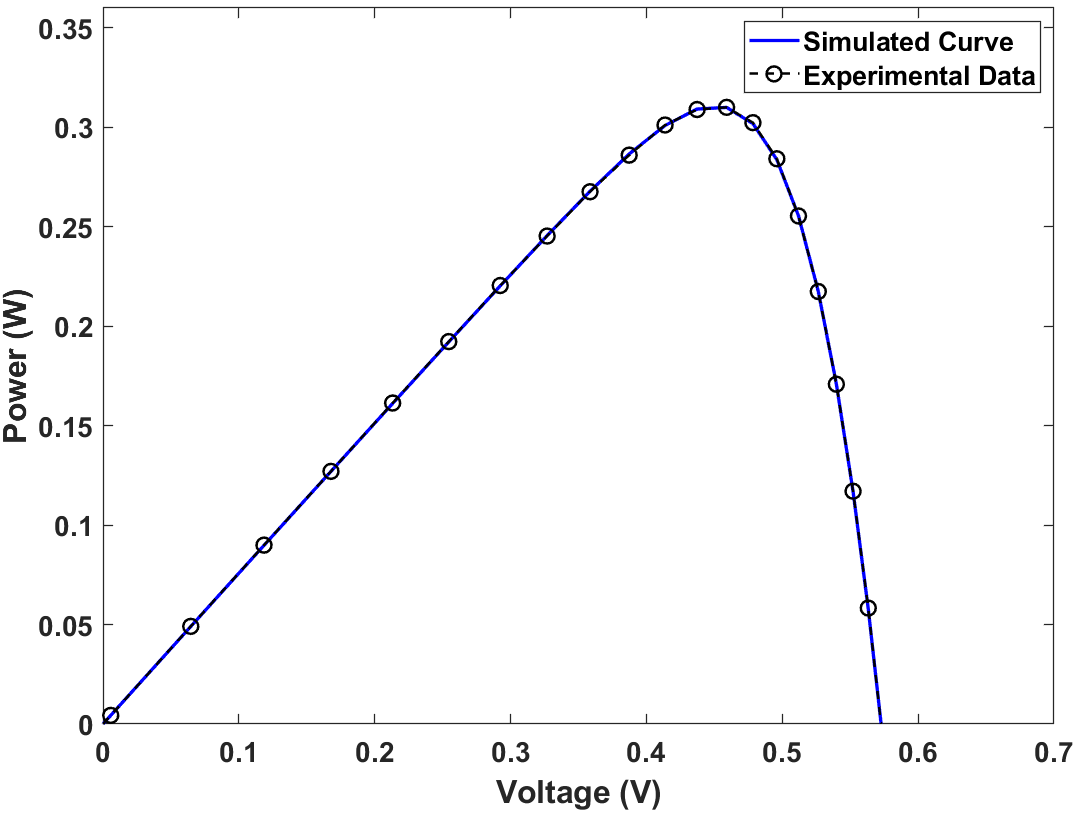}
	 	}%
	 	\captionsetup{justification=raggedright,singlelinecheck=false}	
	 	\caption{Simulated and experimental curves of silicon cell RTC France}
	 	\label{fig:RT_IV}
	 \end{figure}
	 \begin{figure}[!t]
	 	\centering
	 	\includegraphics[width= 3.5 in, height= 2.8 in]{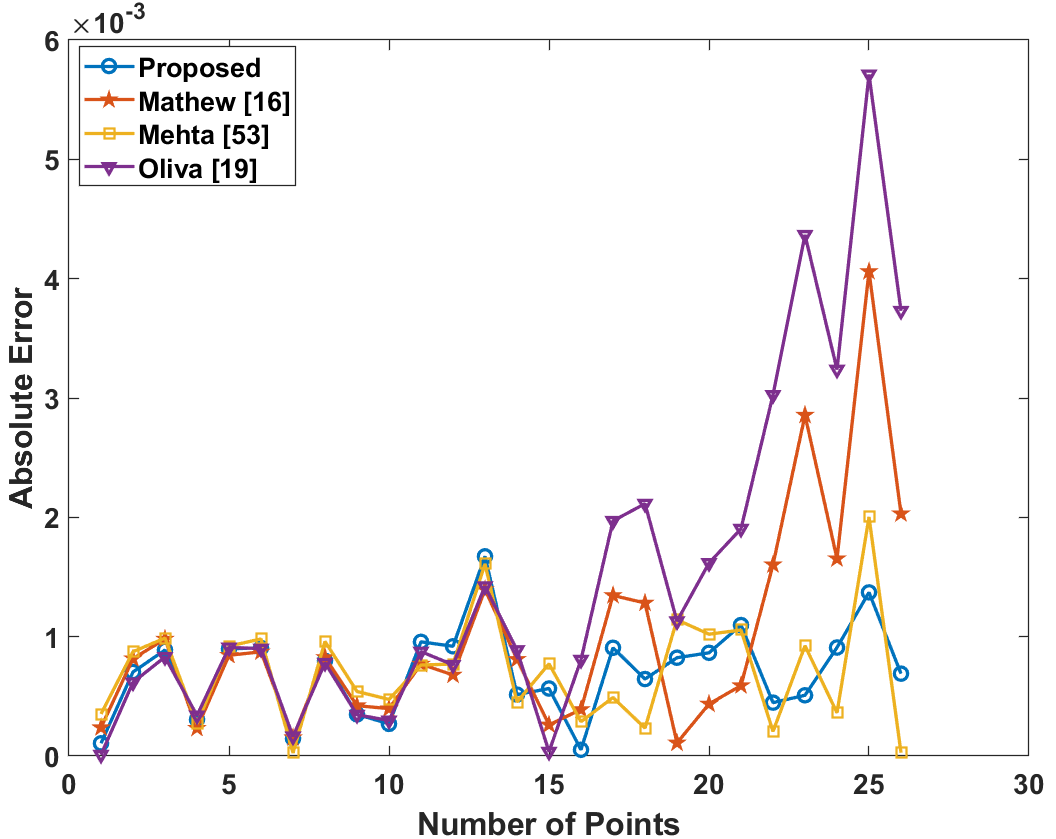}
	 	\captionsetup{justification=raggedright,singlelinecheck=false}
	 	\caption{Comparison of absolute errors for silicon cell RTC France}
	 	\label{fig:Error_RT}
	 \end{figure}

	 \subsection{Proposed Equation}
	 The comprehensive literature review presented above makes it evident that so far there is no publication that offers the fifth equation without any inaccurate assumption or computational complexity. This manuscript presents an exact analytical equation without these drawbacks. The proposed equation employs the fact that the power-current ($P$-$I$) curve of a PV module also exhibits a bell shaped characteristics with unique maxima like the power-voltage ($P$-$V$) curve as shown in Fig. \ref{fig:overview}(c). Hence the fifth equation is formulated as follows.
	 \begin{equation}
	 \left.\frac{dP}{dI}\right\vert_{{V=V_{MPP}}\atop{I=I_{MPP}}}=0	
	 \label{eq:16}
	 \end{equation}
	 After replicating the steps of fourth equation from (\ref{eq:7}) to (\ref{eq:10}), the final form of the (\ref{eq:16}) is
	 \begin{equation}
	 \begin{split}
	 \left.\frac{dP}{dI}\right\vert_{{V=V_{MPP}}\atop{I=I_{MPP}}}= 0 = V_{MPP}-I_{MPP}\ \times\\\frac{1+\frac{I_s.R_s}{N_s.n.v_t}[\exp(\frac{V_{MPP}+I_{MPP}R_s}{N_s.n.v_t})-1]+\frac{R_s}{R_{sh}}}{\frac{I_s}{N_s.n.v_t}[\exp(\frac{V_{MPP}+I_{MPP}R_s}{N_s.n.v_t})-1]+\frac{1}{R_{sh}}}
	 \label{eq:17}
	 \end{split}
	 \end{equation}
	 
	 Finally, the (\ref{eq:3})–(\ref{eq:5}), (\ref{eq:11}) and (\ref{eq:17}) constitute the complete set of simultaneous independent equations for the solution of five unknown parameters of single diode PV model i.e. $I_{ph}$, $I_s$, $n$, $R_s$ and $R_{sh}$ as summarized in Fig. \ref{fig:overview}.
	 \section{Results and Validation}
	 The efficacy of the proposed approach is established in comparison with the measured experimental data of the PV modules of two different makes: 1) PWP-201 having 36 poly-crystalline series connected silicon cells, experimented at 45$ ^{\circ}$C and 1000 W/m$^2$ 2) RTC France silicon cell experimented at 33$ ^{\circ}$C and 1000 W/m$^2$. The experimental data in the form of the $I$-$V$ characteristic curves of these two PV modules was initially reported in \cite{easwarakhanthan1986nonlinear} and later used by many researchers. Table \ref{tab:parameters} shows the datasheet specifications of both the PV sources. Substituting these boundary values in (\ref{eq:3})–(\ref{eq:5}), (\ref{eq:11}) and (\ref{eq:17}), a definite set of five equations is established to solve for the simulation model parameters. The initial guesses of the parameters are determined through the method presented in \cite{ma2019data}. The equations are solved for the unknown parameters \{$I_{ph}$, $I_s$, $n$, $R_s$, $R_{sh}$\} using the $fslove$ function in MATLAB. Finally plugging these parameters in (\ref{eq:1}) yields the simulation model for the particular PV modules.
	 
	 Fig. \ref{fig:PWP_IV} and \ref{fig:RT_IV} shows the experimental and estimated $I$-$V$ and $P$-$V$ characteristic curves for PWP-201 and RTC France PV modules respectively. It is evident from these figures that simulated curves from the proposed method successfully replicate the experimentally measured characteristics. The accuracy of the proposed method in terms of the closeness of the simulated characteristic curve derived from (\ref{eq:1}) to the measured $I$-$V$ curve, is determined through the root mean square error ($RMSE$)
	 \begin{equation}
	 RMSE= \sqrt\frac{\sum_{i=1}^{i=N}((I_{exp}-I_{sim})^2}{N}
	 \label{eq:18}
	 \end{equation}
	  where $N$ is measurement point number, $i$ is the iteration number, $I_{exp}$ is the experimental value and $I_{sim}$ is the simulated value of PV current. The absolute of the difference between $I_{exp}$ and $I_{sim}$ is called absolute error.
	  
	  The proposed approach is benchmarked against the most recent publications to prove its supremacy not only with respect to the single diode models \cite{mehta2019accurate,jadli2017new,kareem2016new,ding2014simplified} but the double diode models \cite{mathew2017wind,oliva2017parameter} as well. The parameters reported in \cite{mehta2019accurate,jadli2017new,mathew2017wind,ding2014simplified,oliva2017parameter,kareem2016new} are converted into their simulated characteristics curves using the MATLAB $lsqnonlin$ function. The absolute error of the proposed technique in contrast with the benchmarked algorithms is plotted in Fig. \ref{fig:Error_PWP201} and \ref{fig:Error_RT}, which clearly demonstrate that the error of the proposed technique mostly remains much lesser than the other's. Conclusively the performance comparison presented 
	  in Table \ref{tab:final}, that enlists the parameters extracted by various techniques and the associated $RMSE$, succinctly establishes that the proposed approach with least error is superior of all the techniques found in literature.
	\section{Conclusions}
	\begin{center}  
	  	\begin{table*}[t!]
	  		\caption{Comparison of Proposed Approach with Existing Approaches for PV Module and Silicon Cell}
	  		\centering
	  		\begin{tabular}{cccccccccc}
	  			\hline \hline
	  			\textbf{PV Source} & \textbf{Technique} & \textbf{$I_{ph}$ (A)} & \textbf{$I_s$/$I_{s1}$ ($\mu$A)} & \textbf{$I_{s2}$ ($\mu$A)}& \textbf{$R_s$($\Omega$)} & \textbf{$R_{sh}$($\Omega$)}& \textbf{$n/n_1$}& \textbf{$n_2$} & \textbf{$RMSE$}\\\\
	  			\hline
	  			PWP-201 & Proposed  & 1.0324101 & 4.58 & - & 1.200576 & 1587.571 & 1.3807 &-& 3.68$\times$ $10^{-3}$  \\
	  			&  \cite{jadli2017new}  & 1.0322 & 1.4586  & - & 1.338 & 616.751 & 1.2652 &-& 3.73$\times$ $10^{-3}$  \\
	  			& \cite{mehta2019accurate}  & 1.033285 & 1.82  & - & 1.357607 & 850.7068 & 1.2857 &-& 5.18$\times$ $10^{-3}$  \\
	  			& \cite{ding2014simplified}  & 1.033769 & 1.11  & - & 1.426154 & 687.5329 & 1.2393 &-& 6.24$\times$ $10^{-3}$  \\
	  			& \cite{kareem2016new}  & 1.033774 & 1.10  & - & 1.432646 & 689.0408 & 1.2391 &-& 6.54$\times$ $10^{-3}$  \\
	  			\hline
	  			RTC Cell & Proposed  & 0.760810 & 32.65 & - & 0.036234 & 54.0092 & 1.4830 &-& 7.97$\times$ $10^{-4}$  \\
	  			& \cite{ding2014simplified}  & 0.76086 & 27.74 & - & 0.03696 & 49.8889 & 1.4664 & - & 8.24$\times$ $10^{-4}$  \\
	  			& \cite{mehta2019accurate}  & 0.760883 & 29.6 & - & 0.036499 & 51.2596 & 1.4731 &-& 8.46$\times$ $10^{-4}$  \\
	  			& \cite{mathew2017wind}  & 0.7609 & 1.466 & 0.257 & 0.0367 & 53.245 & 2.3776 &1.4604& 13.3$\times$ $10^{-4}$  \\
	  			& \cite{oliva2017parameter}  & 0.76077 & 0.241 & 0.60 & 0.03666 & 55.2016 & 1.45651 & 1.9899 & 20.5$\times$ $10^{-4}$  \\
	  			\hline \hline
	  		\end{tabular}%
	  		\label{tab:final}%
	  	\end{table*}%
	  \end{center}
	 A definite set of five independent equations is imperative for parameters extraction of a single diode model of the PV module. While the first four equations can be easily derived from the key data points available in the datasheet, the formulation of fifth equation is the real challenge. The approaches found in literature till date for formulation of the fifth equation broadly lie in two categories. Some rely on complex algorithms with iterative meta-heuristic search for parameters which need in prior the large number of measured $I$-$V$ data and also are slower in computation. The second category is of the techniques striving for the analytical equations; however all of these published till date use assumptions with inherent inaccuracies which eventually lead to disparity in the simulated model with respect to the experimental data. This manuscript proposed an exact analytical approach using the MPP at the $P$-$I$ characteristic curve of the PV module. The proposed method is free of any complexity or inaccurate assumption. Resultantly the proposed method is proven to be the most accurate of all.
	  
	\let\oldthebibliography\thebibliography
	\let\endoldthebibliography\endthebibliography
	\renewenvironment{thebibliography}[1]{
		\begin{oldthebibliography}{#1}
			\setlength{\itemsep}{0em}
			\setlength{\parskip}{0em}
		}
		{
		\end{oldthebibliography}
	}

	\bibliographystyle{IEEEtran} 
	\bibliography{IEEEabrv,Solar}
	\vskip 0pt plus -1.1fil
	
	\vfill
\end{document}